# Effects of the Cu off-stoichiometry on transport properties of wide gap *p*-type semiconductor, layered oxysulfide LaCuSO

Yosuke Goto,[1,a)] Mai Tanaki,[1] Yuki Okusa,[1] Taizo Shibuya,[2] Kenji Yasuoka,[2] Masanori Matoba,[1] and Yoichi Kamihara[1]

[1]Department of Applied Physics and Physico-Informatics, Faculty of Science and Technology, Keio University, 3-14-1 Hiyoshi, Yokohama 223-8522, Japan
[2]Department of Mechanical Engineering, Faculty of Science and Technology, Keio University, Yokohama 223-8522, Japan
[a)]Electronic mail: ygoto@z8.keio.jp

**Abstract**

Layered oxysulfide LaCu$_{1-x}$SO ($x$ = 0–0.03) was prepared to elucidate the effect of Cu off-stoichiometry on their electrical and thermal transport properties. Electrical resistivity drastically decreases down from ~$10^5$ Ωcm to ~$10^{-1}$ Ωcm as a result of Cu deficiency ($x$ = 0.01) at 300 K. Thermal conductivity of the samples at 300 K, which is dominated by lattice components, is estimated to be 2.3(3) Wm$^{-1}$K$^{-1}$. Stoichiometric LaCuSO has an optical band gap of 3.1 eV, while broad optical absorption at photon energies of approximately 2.1 eV was observed for Cu-deficient samples. Density functional theory calculation suggests that these broad absorption structures probably originate from the in-gap states generated by the sulfur vacancies created to compensate the charge imbalance due to Cu off-stoichiometry. These results clearly demonstrate that Cu deficiency plays a crucial role in determining the electrical transport properties of Cu-based *p*-type transparent semiconductors.

There has been considerable interest in the study of the electronic phases of mixed anion compounds with a ZrCuSiAs-type structure, such as LnT$_M$PnO and LnT$_M$ChO (Ln = lanthanide, T$_M$ = transition metal, Pn = pnictogen, Ch = chalcogen), owing to their unique and exotic functionalities including transparent *p*-type conductivity,[1–4] iron-based superconductivity,[5–8] and thermoelectricity.[9,10] These compounds belong to the tetragonal *P*4/*nmm* space group with the crystal being a layered structure composed of an alternate stack of LnO and (T$_M$Ch, T$_M$Pn) layers.[11,12] The carrier conduction path of the T$_M$Ch layer is sandwiched by the wide band gap LnO layers, which may be regarded as a multiple-quantum well embedded in the crystal structure.[13]

A typical example of the ZrCuSiAs-type compounds is the oxysulfide LaCuSO, which exhibits *p*-type electrical conductivity with a band gap of 3.1 eV.[1] Furthermore, LaCuSO exhibits blue to near-ultraviolet light emission[14–17] and large third-order optical nonlinearity[18] because of the presence of room-temperature stable excitons. The electrical transport properties of LaCuSO are extremely sensitive to its chemical composition, as summarized in Table S1.[19] For example, the electrical resistivity of "undoped" LaCuSO polycrystals ranges from ~$10^1$ Ωcm to ~$10^5$ Ωcm,[15,20–23] whereas the carrier concentration can be controlled in the range $10^{15}$–$10^{20}$ cm$^{-3}$ by the substitution of $Sr^{2+}$ ions for $La^{3+}$.[3] Although the $Sr^{2+}$ substitution is considered favorable for $p$-type carrier doping, the actual origin of $p$-type conduction was reported to be due to conducting holes generated by Cu off-stoichiometry.[24–26] This can be explained based on the fact that the valence band is composed of the hybridization between Cu $3d$ orbitals and S $2p$ orbitals[27,28] as well as other Cu-based oxides, including $Cu_2O$ or $CuAlO_2$.[29–32] Furthermore, a detailed study of the LaCuSeO and BaCuChF substantiated the fact that Cu deficiency contribute to degenerate conduction.[26,33]

These reports strongly suggest that Cu vacancies act as acceptor-like defect in the alkali earth metal ion (Ae)-doped LaCuSO. However, the transport properties of Ae-free LaCuSO containing Cu deficiency have not been established so far. In this study, we have prepared polycrystalline LaCuSO with and without deficiency to elucidate the effect of Cu off-stoichiometry on their electrical and thermal transport properties.

Polycrystalline samples were prepared by the solid-state reaction of the precursors in a sealed silica tube. In the typical process, commercial $La_2O_3$ powder (Kojundo Chemical, 99.99%) was dehydrated at 600 °C for 10 h in air. Following that, the mixture containing La (Nippon Yttrium, 99.9%), Cu (Kojundo Chemical, 99.9%), and S (Kojundo Chemical, 99.99%) powders in the stoichiometric ratio of 1:3:3 (133 powder) was heated at 800 °C for 10 h in an evacuated silica tube. Subsequently, a 1:1 mixture of the 133 powder with dehydrated $La_2O_3$ was pressed and heated in a sealed silica tube at 1000 °C for 40 h to obtain a sintered pellet. All the above mentioned procedures were performed in an Ar-filled glove box (MIWA Mfg; $O_2$, $H_2O$ < 1 ppm). Furthermore, to obtain Cu-deficient LaCuSO, we prepared starting materials of CuS, a mixture of La and S in a stoichiometric ratio of 1:3 (13 powder), and $La_2S_3$ powder. These starting materials were obtained by solid-state reaction in a sealed silica tube at 400 °C for CuS and 13 powder, and at 1000 °C for $La_2S_3$, respectively. The relative density of the sintered pellets was estimated to be approximately 65%.

Purity of the obtained samples was examined by X-ray diffraction (XRD) using CuKα radiation (Rigaku Rint 2500). Hall effect measurements were performed by five-probe geometry under a magnetic field from −0.43 T to 0.43 T. Electrical resistivity ($\rho$) was measured by the four-probe technique using Ag paste as the electrode. Seebeck coefficient ($S$) was measured by copper–constantan thermocouples using strain gauge as the heater. Thermal conductivity ($\kappa$) was obtained by the heat flux from the strain gauge and the generated temperature difference under the pressure of $10^{-3}$ Pa. Total diffuse reflectance spectra was measured using a spectrometer equipped with an integrating sphere (Hitachi High-Tech, U-4100). $Al_2O_3$ powders were used as the standard reference. Optical absorption coefficient ($\alpha$) was determined from the reflectivity ($R$) value using the following Kubelka–Munk equation,[34] $(1 − R)^2/2R = \alpha/s$, where $s$ denotes the scattering factor.

Theoretical band structures within density functional theory (DFT) were computationally simulated using the plane-wave projector augmented-wave[35,36] method implemented in the VASP code.[37,38] The exchange-correlation potential was treated within generalized gradient approximation by the Perdew–Becke–Ernzerhof method.[39] A $3 \times 3 \times 1$ superlattice cell was constructed to investigate the defect-containing LaCuSO, namely, $La_{17}Cu_{18}S_{18}O_{18}$, $La_{18}Cu_{17}S_{18}O_{18}$, $La_{18}Cu_{18}S_{17}O_{18}$, $La_{18}Cu_{18}S_{18}O_{17}$. The Brillouin zone was sampled by a $2 \times 2 \times 3$ Monkhorst–Pack grid[40] and a cutoff of 450 eV was chosen for the plane-wave basis set. Density of states (DOS) was calculated using the relaxed structures, where the Hellman–Feynman forces were reduced until 0.1 eVnm$^{-1}$.

Figure 1 shows the XRD patterns of $LaCu_{1-x}SO$ ($x$ = 0–0.03). Almost all the diffraction peaks could be assigned to those of the tetragonal phase, indicating that ZrCuSiAs-type structure is the dominant phase in these samples. Although some diffractions peaks corresponding to the impurity phases can be observed, the amounts of these impurities are observed to be lesser than 1 mol%. The lattice constants of LaCuSO with Cu deficiency were smaller than that of stoichiometric LaCuSO at approximately 0.1%, with $x$ = 0.02 sample showing the smallest lattice volume.

Figure 2(a) shows the $\rho$ as a function of $T$. The $\rho$ of LaCuSO ($x$ = 0) is $6.7 \times 10^5$ $\Omega$cm at 300 K, and it increases with decreasing temperature with an activation energy of 0.28 eV. The $\rho$ drastically decreases to 0.26 $\Omega$cm for $x$ = 0.01 at 300 K as a result of Cu deficiency. The $\rho$–$T$ curve of the $LaCu_{1-x}SO$ ($x$ = 0.01–0.03) samples is almost independent of $T$, although the temperature coefficient (d$\rho$/d$T$) is slightly negative in contrast to the positive temperature coefficient of Sr-doped or Ca and Ni co-doped LaCuSO.[21,41] Hall voltage of samples are hindered by thermal fluctuation of parasitic ohmic voltage due to mislocation of electrodes. Hall coefficient of LaCuSO with Cu deficiency was less than $10^{-6}$ mC$^{-1}$, suggesting carrier concentration was higher than $10^{19}$cm$^{-3}$. The carrier mobility of LaCuSO with Cu off-stoichiometry was estimated to be less than 1 cm$^2$V$^{-1}$s$^{-1}$. The p-type conductivity of the samples was confirmed from the positive value of the $S$, as shown in Figure 2(b). The value of $S$ was approximately 20 $\mu$VK$^{-1}$ for $x$ = 0.01–0.03 at 300 K, which is higher than those of the Sr-doped polycrystalline bulk sample.[4]

Figure 2(c) shows $\kappa$ of the samples as a function of $T$. Results indicate that the value of $\kappa$ is 2.3(3) Wm$^{-1}$K$^{-1}$ at 300 K. Furthermore, it increases with decreasing $T$, indicating that the thermal transport at these temperatures is predominantly via phonon–phonon Umklapp scattering.[42] In a simple kinetic picture, $\kappa$ is expressed as the sum of the electrical component ($\kappa_{el}$) and lattice component ($\kappa_L$), as $\kappa = \kappa_{el} + \kappa_L$. Here $\kappa_{el}$ is approximately described as $LT\rho^{-1}$, where $L$ is the Lorentz number, being $L \sim 2.4 \times 10^{-8}$ V$^2$K$^{-2}$ for metal or degenerate semiconductor. In the present study, the value of $\kappa_{el}$ was evaluated to be $2.8 \times 10^{-3}$ Wm$^{-1}$K$^{-1}$ for $x$ = 0.01, which is less than 1% of the total $\kappa$. This indicates that the lattice component dominates the thermal transport properties of the samples. Given the fact that the $\kappa$ value of LaCuSO has not been reported so far in the literature, the value was compared with that of the corresponding isostructural compounds such as LaCuSeO and BiCuSeO.[43,44] Intriguingly, the $\kappa$ value of LaCuSO was observed to be almost similar to that of LaCuSeO.[43] This is rather interesting, as in many compounds the presence of heavier chalcogen (Se) atoms tends to reduce the $\kappa$ value because of lower sound velocity. The relatively low value of $\kappa$ observed in this study probably originates from the large Grüneisen parameter and low Young's modulus of the layered crystal structure with weak bonding between the LaO and CuS layers.[44] Nevertheless, the $\kappa$ value of LaCuSO is still higher than those of BiCuSeO ($\kappa \sim 1$ Wm$^{-1}$K$^{-1}$ at 300 K),[9] wherein the lone pair electrons in Bi$^{3+}$ with an electron configuration of $d^{10}s^2p^0$ distort the BiO layers and correspondingly reduce the $\kappa$ value.[45] It should be noted that dense sample is required to discuss the $\kappa$ quantitatively.

Figure 3(a) shows the reflectance spectra of LaCu$_{1-x}$SO. The stoichiometric LaCuSO was a brownish white in color, whereas the LaCuSO with Cu deficiency was a dark-colored product, as shown in the inset of Figure 3(a). The observed change in color is clearly observed in the reflectance spectra. The $x = 0$ sample exhibited high reflectivity in the visible region. On the other hand, the reflectivity of $x = 0.01$–$0.03$ samples decreased to approximately 0.1. This could be attributed to the light absorption due to the sulfur defect states, as discussed later. Figure 3(b) shows $(\alpha h \nu s^{-1})^2$ converted from reflectivity plotted as a function of the photon energy. In case of undoped LaCuSO, results indicate direct-type transition at 3.1 eV arising from the room-temperature stable excitons.[14] On the other hand, in case of the Cu-deficient samples, we could observe a broad absorption structure at approximately 2.1 eV.

To examine the possible defects we estimated the formation energy ($E_{form}$) of neutral point defects using DFT calculation. For example, the $E_{form}$ of Cu vacancies (V$_{Cu}$) was estimated based on the chemical reactions of (LaCuSO)$_{18}$ → La$_{18}$Cu$_{17}$S$_{18}$O$_{18}$ + Cu (fcc crystal). As shown in Table 1, the $E_{form}$ of V$_{Cu}$ was close to zero, whereas those for other vacancies were larger than 2 eV, indicating that V$_{Cu}$ is the most easily created in LaCuSO. Additionally, the $E_{form}$ of [2V$_{Cu}$ + V$_S$] complex defects was smaller than that of V$_S$ monovacancies, suggesting sulfur vacancies were created to compensate the charge imbalance due to Cu vacancies. Although $E_{form}$ of [2V$_{Cu}$ + V$_S$] complex defects was larger than energy width of thermal fluctuations (~0.11 eV at temperatures for sample synthesis), we note that both positive and negative values for $E_{form}$ of [2V$_{Cu}$ + V$_{Se}$] was reported for LaCuSeO.[26,46] Generally, $E_{form}$ depends on the Fermi energy and the charge of defects. In the case of BaCuSF, $E_{form}$ of neutral V$_S$ is larger than 2 eV, while that of V$_S^{2+}$ was close to zero.[33] Although several "provided" charged defects enable us to discuss stability of complex defects, the detailed theoretical study of defect physics for LaCuSO is beyond our scope. Especially, the defect transition of V$_{Cu}$ was suggested as the origin of subgap absorption of LaCuSeO.[46]

Figure 4 shows the theoretical DOS of LaCuSO with and without vacancies. The V$_{Cu}$ generate additional states in the vicinity of the valence band maximum (VBM) of the stoichiometric LaCuSO. On the other hand, the V$_S$ generate the donor-like states below the conduction band minimum (CBM), which is similar to the case of LaCuSeO.[26] Although underestimation of the band gap for DFT does not allow us to quantitatively elucidate the origin of broad absorption, it is probably due to the in-gap states generated by the V$_s$ created to compensate the V$_{Cu}$ in LaCuSO with Cu off-stoichiometry. We note that band gap of LaCuSO obtained by DFT calculation, which is determined by energy difference between VBM and CBM, was 1.7 eV and in 55% of an optical band gap of 3.1 eV via reflectivity measurements. The in-gap states created by sulfur defects in the [2V$_{Cu}$ + V$_S$] model locate 1.0 eV above VBM, providing 1.8 eV when we shifted assuming a fixed rate of 55% for theory to experiments. This value is in reasonable agreement with additional absorption structure at approximately 2.1 eV observed in LaCuSO with Cu deficiency.

In summary, this study elucidates the effect of Cu off-stoichiometry on the transport properties of LaCuSO. The presence of Cu deficiency results in a drastic decrease in the electrical resistivity from ~$10^5$ to ~$10^{-1}$ Ωcm. The thermal conductivity of LaCu$_{1-x}$SO, which is dominated by lattice components, is estimated to be 2.3(3) Wm$^{-1}$K$^{-1}$ at 300 K. The Cu-deficient samples exhibit broad absorption at photon energies of approximately 2.1 eV. DFT calculation suggests these broad absorption structures probably due to the in-gap states generated by the sulfur deficiency created to compensate the Cu off-stoichiometry in LaCuSO. These results clearly indicate that the Cu deficiency have a crucial impact on the electrical transport properties of a Cu-based *p*-type semiconductor.

**Acknowledgements**

This work was partially supported by the research grants from Keio University, the Keio Leading-edge Laboratory of Science and Technology (KLL), Funding Programs for World-Leading Innovative R&D on Science and Technology (FIRST from Japan Society for Promotion of Science (JSPS)), and Hitachi Metals Materials Science Foundation (HMMSF).

**Table caption**

Table 1.

Formation energy ($E_{\text{form}}$) of neutral point defects and complex defects in LaCuSO.

**Figure captions**

Figure 1.

X-ray diffraction patterns of LaCu$_{1-x}$SO. Diffraction intensity is plotted on a logarithmic scale. Vertical bars at the bottom represent calculated angles of Bragg diffractions for LaCuSO. Arrows represent Bragg diffractions because of secondary phases. Lattice constants (*a* and *c*) are *a* = 0.399927(5) nm and *c* = 0.85216(2) nm for *x* = 0, *a* = 0.39961(1) nm and *c* = 0.85136(3) nm for *x* = 0.01, *a* = 0.39956(1) nm and *c* = 0.85115(2) nm for *x* = 0.02, *a* = 0.39956(3) nm and *c* = 0.85143(1) nm for *x* = 0.03. The values in parentheses for *a* and *c* are the statistical errors. Other errors such as temperature fluctuations (<1 K) should be considered.

Figure 2.

Electrical and thermal transport properties versus temperature (*T*) of LaCu$_{1-x}$SO. (a) Electrical resistivity ($\rho$), (b) Seebeck coefficient (*S*), and (c) thermal conductivity ($\kappa$).

Figure 3.

(a) Reflectivity spectra of LaCu$_{1-x}$SO. The inset shows photographs of sintered sample. (b) The ($\alpha h \nu s^{-1}$)$^2$ versus $h\nu$ curve converted from reflectivity spectra. The black arrow denotes absorption edge at 3.1 eV, while red arrow of *x* = 0.01 represent additional absorption edge of LaCuSO with Cu deficiency.

Figure 4.

Theoretical density of states (DOS) for stoichiometric and defect-containing LaCuSO. The vertical dashed lines denote the Fermi level. The black arrows represent the electronic states generated by vacancies. Energy is aligned by La 5*s* states at around −32 eV.

Table 1

| defect model | $V_{La}$ | $V_{Cu}$ | $V_S$ | $V_O$ | $2V_{Cu} + V_S$ |
|---|---|---|---|---|---|
| $E_{form}$ (eV) | 8.15 | 0.19 | 2.63 | 6.62 | 2.17 |

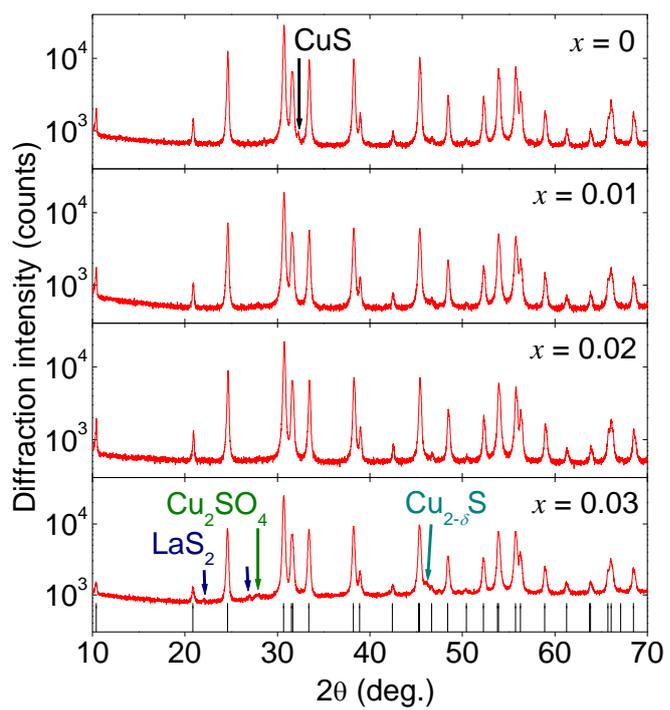

Figure 1

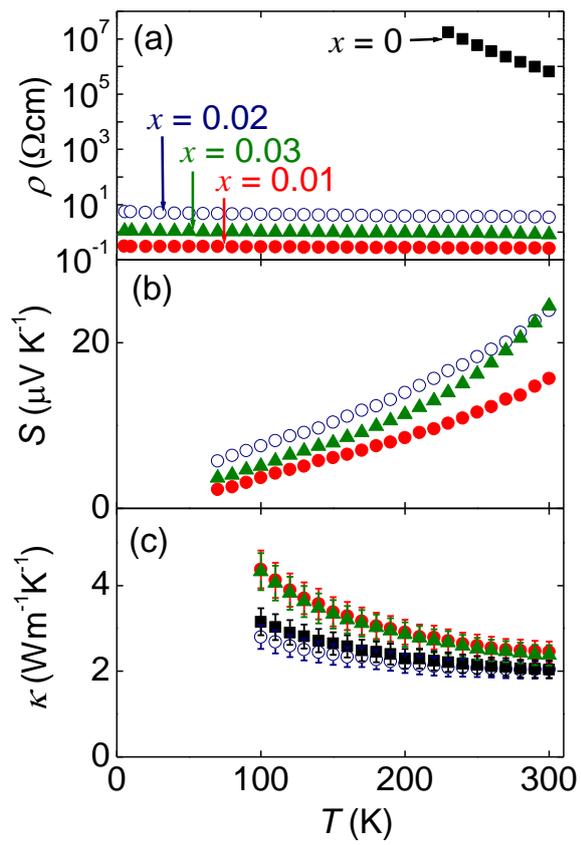

Figure 2

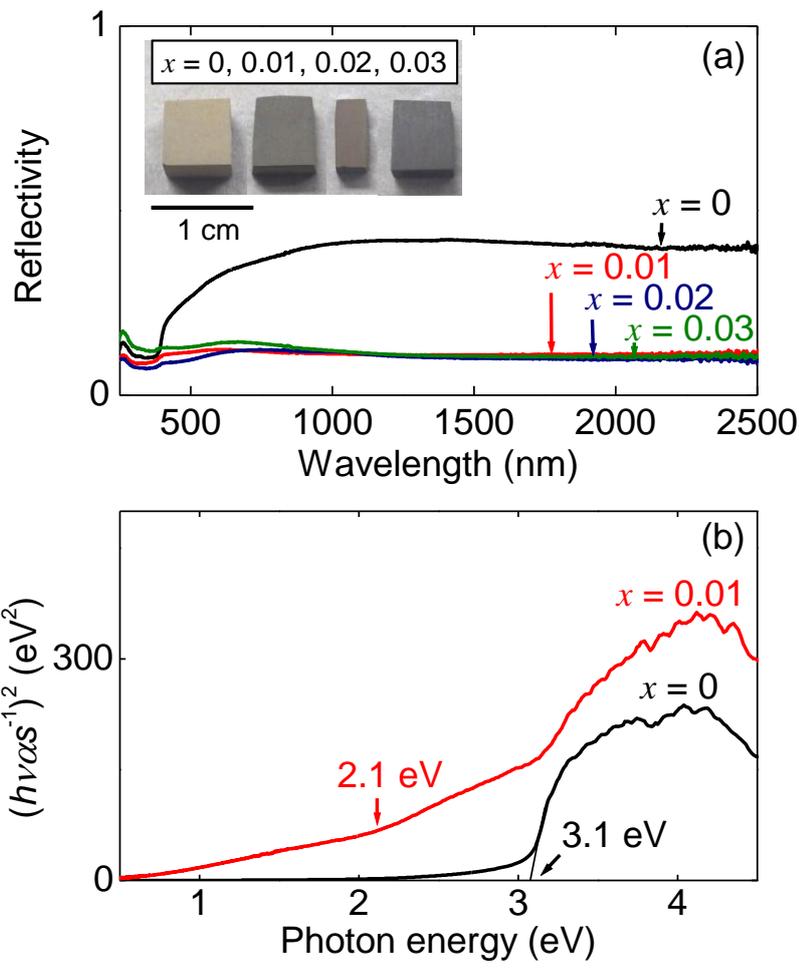

Figure 3

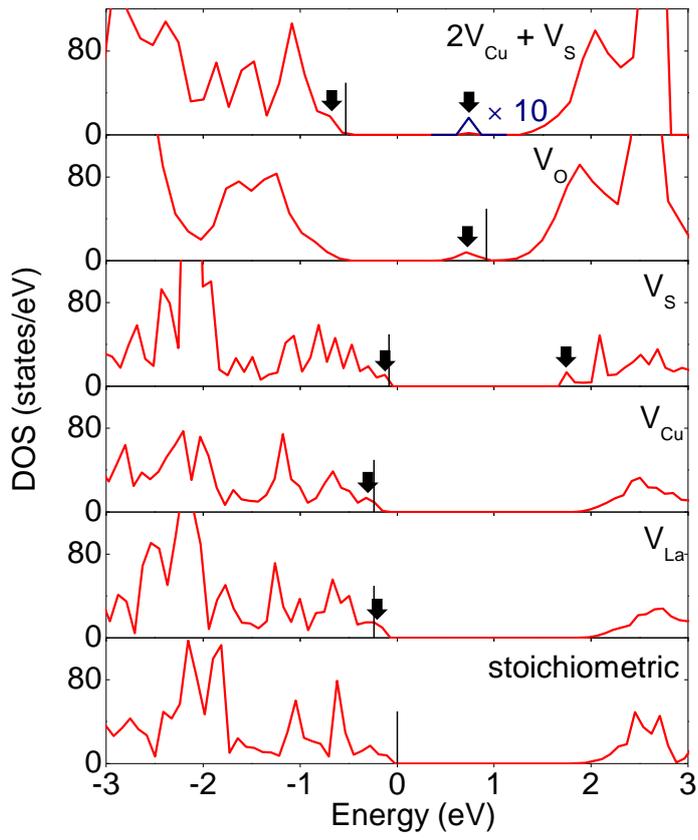

Figure 4

Supporting information for "Effects of the Cu off-stoichiometry on transport properties of wide gap *p*-type semiconductor, layered oxysulfide LaCuSO"

Yosuke Goto,[1] Mai Tanaki,[1] Yuki Okusa,[1] Taizo Shibuya,[2] Kenji Yasuoka,[2] Masanori Matoba,[1] and Yoichi Kamihara[1]

[1]Department of Applied Physics and Physico-Informatics, Faculty of Science and Technology, Keio University, 3-14-1 Hiyoshi, Yokohama 223-8522, Japan
[2]Department of Mechanical Engineering, Faculty of Science and Technology, Keio University, Yokohama 223-8522, Japan

Table S1. Electrical transport properties of LaCuSO at 300 K.[a]

| sample | | $\rho$ ($\Omega$cm) | $S$ ($\mu$VK$^{-1}$) | $n_H$ (cm$^{-3}$) | $\mu_H$ (cm$^2$V$^{-1}$s$^{-1}$) | reference |
|---|---|---|---|---|---|---|
| LaCuSO | polycrystalline bulk | $1.0 \times 10^5$ | — | — | — | [1] |
| | | $2.0 \times 10^1$ | — | — | — | [2] |
| | | $1.1 \times 10^{2}$,[b] | 89,[b] | | | [3] |
| | | $1.3 \times 10^{1}$,[c] | 50,[c] | — | — | |
| | | $(1.0 \times 10^1)$[d] | 20[d] | | | |
| | | $1.3 \times 10^5$ | 150 | — | — | [4] |
| | | $1.8 \times 10^5$ | — | — | — | [5] |
| | | $7.5 \times 10^1$ | 134 | $3.0 \times 10^{15}$ | 9.4 | [6] |
| | | $6.7 \times 10^5$ | — | $(4.7 \times 10^{13})$[e] | $(2.0 \times 10^{-1})$[e] | present study |
| | polycrystalline thin film | $8.3 \times 10^1$ | 150 | — | — | [7] |
| | | $1.6 \times 10^4$ | 713 | $(2.0 \times 10^{15})$[e] | $(2.0 \times 10^{-1})$[e] | [8] |
| | epitaxial thin film | ~$10^5$ | — | — | — | [9] |
| | | $1.5 \times 10^0$ | 122 | $1.0 \times 10^{19}$ | $5.0 \times 10^{-1}$ | [10] |
| LaCu$_{1-\delta}$SO | polycrystalline bulk | $8.0 \times 10^{3}$,[f] | — | — | — | [5] |
| | | $(7.5 \times 10^1)$[g] | | | | |
| LaCu$_{0.99}$SO | | $2.6 \times 10^{-1}$ | 16 | > $10^{19}$ | < $10^0$ | present study |
| LaCu$_{0.98}$SO | | $3.5 \times 10^0$ | 24 | > $10^{19}$ | < $10^0$ | present study |
| LaCu$_{0.97}$SO | | $8.2 \times 10^{-1}$ | 24 | > $10^{19}$ | < $10^0$ | present study |
| Sr-doped LaCuSO | | $(5.3 \times 10^0)$[h] | 3 | — | — | [4] |
| | | $(7.5 \times 10^{-2})$[i] | — | — | — | [11] |
| La$_{0.9}$Ca$_{0.1}$Cu$_{0.9}$Ni$_{0.1}$SO | | $1.2 \times 10^{-1}$ | — | — | — | [2] |
| La$_{0.97}$Sr$_{0.03}$CuSO | polycrystalline thin film | $5.0 \times 10^{-2}$ | 44 | $2.7 \times 10^{20}$ | $4.7 \times 10^{-1}$ | [8] |
| Sr-doped LaCuSO[g] | polycrystalline thin film | $3.8 \times 10^0$ | 40 | — | — | [7] |

[a] Electrical resistivity ($\rho$), Seebeck coefficient ($S$), Hall carrier concentration ($n_H$), and Hall mobility ($\mu_H$).

[b] Sintered at 800 °C for 6 h.

[c] Sintered at 900 °C for 6 h.

[d] Sintered at 900 °C for 40 h.

[e] The $n_H$ was estimated assuming $\mu_H$ is 0.2 cm$^2$V$^{-1}$s$^{-1}$.

[f] Nominal Cu deficiency is 0.5 mol%.

[g] Nominal Cu excess is 1.5 mol%.

[h] Nominal Sr content is 5 mol%.

[i] Nominal Sr content is 30 mol%.